\pdfoutput=1 

\documentclass[sigconf,authorversion,nonacm]{acmart}

\AtBeginDocument{%
  \providecommand\BibTeX{{%
    \normalfont B\kern-0.5em{\scshape i\kern-0.25em b}\kern-0.8em\TeX}}}

\acmConference[KDD ADS'23]{29th ACM SIGKDD Conference on Knowledge Discovery and Data Mining}{August 06--15,
  2023}{Long Beach, CA}
\acmPrice{15.00}
\acmISBN{978-1-4503-XXXX-X/18/06}

\usepackage{comment}
\usepackage{booktabs} 
\usepackage{multirow}

\usepackage{graphicx}
\graphicspath{ {resource/figure/} }

\usepackage{mathtools}

\makeatletter
\def\@copyrightspace{\relax}
\makeatother

\begin{document}

\title{TPDR: A Novel Two-Step Transformer-based Product and Class Description Match and Retrieval Method}

\author{Washington Cunha}
\authornote{Both authors contributed equally to this research.}
\affiliation{%
  \institution{Federal University of Minas Gerais}
  \country{Brazil}
}
\email{washingtoncunha@dcc.ufmg.br}

\author{Celso França}
\authornotemark[1]
\affiliation{%
  \institution{Federal University of Minas Gerais}
  \country{Brazil}
}
\email{celsofranca@dcc.ufmg.br}

\author{Leonardo Rocha}
\affiliation{%
    \institution{Federal University of São João del Rei}
    \country{Brazil}
}
\email{lcrocha@ufsj.edu.br}

\author{Marcos André Gonçalves}
\affiliation{%
  \institution{Federal University of Minas Gerais}
  \country{Brazil}
}
\email{mgoncalv@dcc.ufmg.br}

\renewcommand{\shortauthors}{Washington Cunha, Celso França, Leonardo Rocha, \& Marcos André Gonçalves}
\newcommand{\red}[1]{\textcolor{red}{#1}}

\begin{abstract}

There is a niche of companies responsible for intermediating the purchase of large batches of varied products for other companies, for which the main challenge is to perform product description standardization, i.e., matching an item described by a client with a product described in a catalog. The problem is complex since the client's product description may be: (1) potentially noisy; (2) short and uninformative;
(e.g., missing information about model and size); 
and (3) cross-language. In this paper, we formalize this problem as a ranking task: given an initial client product specification (query), return the most appropriate standardized descriptions (response). Traditional ranking strategies are difficult to be applied to this problem since there is a single relevant catalog item for every client description. Traditional IR techniques based only on syntactic matching are not enough since the initial client specification (IS) may be very different from the standardized description (SD). Furthermore, syntactic matchings usually cannot focus on the essential parts of the product descriptions and cannot exploit the correlations among these parts. Thus, we propose \textbf{TPDR}, a two-step {\bf T}ransformer-based {\bf P}roduct and Class {\bf D}escription {\bf R}etrieval method that is able to explore the semantic correspondence between IS and SD, by exploiting attention mechanisms and contrastive learning. First, \textbf{TPDR} employs the transformers as two encoders sharing the embedding vector space: one for encoding the IS and another for the SD, in which corresponding pairs (IS, SD) must be close in the vector space. Closeness is further enforced by a contrastive learning mechanism leveraging a specialized loss function. \textbf{TPDR} also exploits a (second) re-ranking step based on syntactic features that are very important for the exact matching (model, dimension) of certain products that may have been neglected by the transformers. To evaluate our proposal, we consider 11 datasets from a real company, covering different application contexts. Our solution was able to retrieve the correct standardized product before the $5^{th}$ ranking position in \textbf{71\%} of the cases and its correct category in the first position in \textbf{80\%} of the situations. Moreover, the effectiveness gains over purely syntactic or semantic baselines reach up to 3.7 times, solving cases that none of the approaches in isolation can do by themselves.\looseness=-1

\end{abstract}

\vspace{-0.3cm}

\maketitle

\section{Introduction}

Large companies that need to buy large batches of varied products commonly hire intermediary companies, such as Astrein\footnote{\url{https://www.astrein.com.br/}} -- a highlight of the branch in Brazil, to perform a product description standardization process in order to avoid delays and financial losses in case of purchase of wrong products. This standardization process consists of  matching, potentially noisy and imprecise client product descriptions (initial specification - IS) lists with fine-grained and standardized detailed product descriptions (standardized description - SD) that precisely describe the desired list of purchases, usually compiled in product catalogs.\looseness=-1

To better illustrate the problem, let us consider the following example scenario to understand how these ``matching'' companies operate. A construction company \textit{X} will build a research complex. Company \textit{X} will be responsible for constructing the complex's buildings and assembling all its laboratories. This company will need an extensive diversity of products, from construction materials such as cement, bricks, structural metals, and screws to office supplies and, potentially, specific research laboratory materials, such as computers with specialized hardware, for example. To make the execution of the project feasible, company \textit{X} obtains from its employees' several purchase lists. These lists are compiled by company \textit{X}´s managers and employees, who will ultimately perform their research activities in the new facilities. The company \textit{X} hires a company \textit{Y} (specialized in intermediating purchases), which, in turn, has the product catalogs of several potential suppliers of the necessary materials. Before placing the orders themselves, company \textit{Y} needs to standardize all the specified items, and only after this process is finalized, it consults the materials with the suppliers. Finally,  company \textit{Y} chooses the best suppliers in order to reduce the total cost and delivery time of purchases. This intermediation is advantageous since other companies can also present similar orders, allowing the purchase in large batches and obtaining product discounts (cost reduction) and special delivery conditions (time reduction). On the other hand, a large purchase of wrong products due to inconsistencies and ambiguities in the initial (client) product list may cause all types of delays and financial problems. \looseness=-1

The most considerable technical challenge in this practical problem is the \textit{vocabulary mismatch} between the usually noisy, imprecise, incomplete, and/or ambiguous initial (client) specification (IS) and the standardized description (SD) detailed in a product's catalog, meaning that the matching task is not an easy one. This challenge refers to a common phenomenon of human communication in which clients who write the description of the product to be purchased and people who built the standardized catalogs of product descriptions likely used different or inconsistent vocabularies with different levels of details. We further elaborate on this problem next.\looseness=-1

\paragraph*{\textbf{Industrial Problem Definition:}} The main challenge for these purchasing intermediation companies is the product description standardization stage. Let us consider the description $IS-X_i$ made by a customer, for example, company $X$. Company $Y$ must be able to retrieve the standardized description $SD-X_i$ from its product catalog, consisting of thousands or even millions of descriptions. In general, customer purchase product specifications are built by putting together several sub-lists that, in turn, are written by people with different backgrounds, with varying levels (or lack thereof) of detail, creating several challenges to be overcome, among them: (1) potentially noisy data; (2) short and uninformative descriptions (missing information about model and size, for example); and (3) multi-language descriptions. We illustrate this problem in Table~\ref{tab:intro_examples} with real examples experienced by a purchases intermediation company.\looseness=-1


\vspace{-0.7cm}
\begin{table}[!h]
    \centering
        \begin{figure}[H]
        	\centering

        	\includegraphics[width=0.46\textwidth]{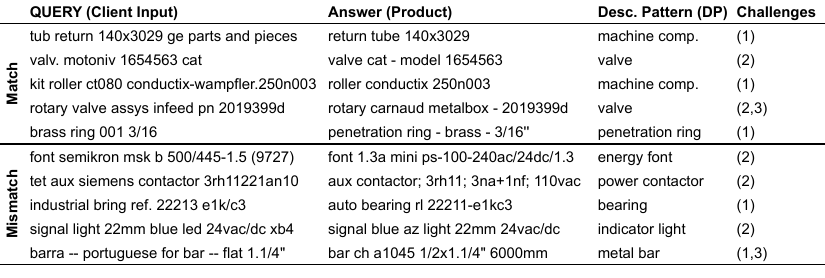}
        \end{figure}
    \vspace{-0.5cm}
    \caption{\small{Examples of the product description standardization.}}
    \label{tab:intro_examples}
\end{table}
\vspace{-0.3cm}
\vspace{-0.7cm}

\paragraph*{\textbf{Failure Implications:}} Minor errors in retrieving standardized items can lead to significant financial losses and severe accounting errors. Errors in the purchasing process lead to a decrease in the agility of the process as a whole. When a product is delivered and, during the inspection, it turns out to be a wrong product, the product exchange is estimated to be
at least \textbf{three} times more time-consuming than the standard delivery process (two deliveries and one return process), according to Astrein. In addition to being a total waste of resources, it can take a long time to resolve it appropriately.\looseness=-1 

\paragraph*{\textbf{Main Goal:}} The matching process we described above can be done manually searching for \textit{keywords}, in a very time-consuming and still prone-to-error process. Our goal here is to help to automate the process as much as possible by addressing two distinct issues in an end-to-end approach: (i) association of customer-supplied material description --\textbf{IS}-- to a previously annotated Description Pattern (\textbf{DP})\footnote{A Description Pattern is equivalent to a category or class of similar products, e.g., a metal tube or valve}, shown in the third column of Table 1; and (ii) Matching the client description with a previously standardized item --\textbf{SD}-- in a product catalog.\looseness=-1
  
\paragraph*{\textbf{Summary of our proposal:}} 
Our proposed solution to the product standardization problem faces the problem as an Information Retrieval task, more explicitly, a ranking task: given an initial specification (query), return, from the product catalog, the most appropriate standardized descriptions (response), as well as their associated DPs. Accordingly, in this paper, we propose \textbf{TPDR} -- a novel {\bf T}ransformer-based {\bf P}roduct and Class {\bf D}escription {\bf R}etrieval  method.  Traditional ranking strategies are difficult to be applied  to this problem since there is a single relevant SD for every single IS, if any exists (the product may not have been  previously cataloged).  Applying traditional IR techniques based on the syntactic descriptions of the products is also not enough, as a customer can present a product description very different from the SD. For example, in Table~\ref{tab:intro_examples}, we have the example IS as ``brass ring 001 3/16" and the corresponding SD is ``penetration ring - brass - 3/16'' ".  Moreover, syntactic matchings are not able to focus on the most important parts of the product descriptions (which may be different depending on the product). They cannot exploit important correlations of these parts (e.g., product type, material, and size). In this sense, it is necessary to adopt ranking strategies that also considerate the \textit{semantic correspondence} between ISs and SDs, as our proposal.\looseness=-1 

\begin{figure}[!htb]
	\centering
	\includegraphics[width=.25\textwidth]{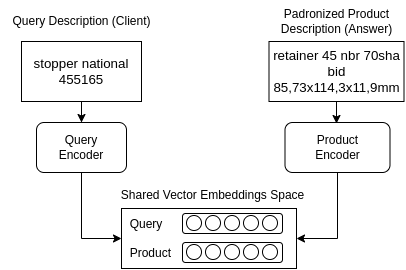}
 \vspace{-0.3cm}
	\caption{Proposed Model}
	\label{fig:proposedmodel}
\end{figure}

Accordingly, \textbf{TPDR}'s first step is to employ transformers as two encoders sharing the embedding vector space: one for encoding the initial specification (query) and another for the standardized descriptions (response).  The transformers exploit \textit{attention} as well as \textit{constrastive learning} mechanisms that allow focusing on the most important parts of both product descriptions (IS and SD) for the sake of the matching as well as relevant correlations among the products´ parts. Figure \ref{fig:proposedmodel} represents this stage of the proposed model. Given a pair (query, response), we aim for both to be close in the shared vector space. In the example of Figure \ref{fig:proposedmodel}, given the initial specification  \textit{``stopper national 455165''} and the standardized description  \textit{``retainer 45 nbr 70sha bid 85,73x114x3x11,9mm''}, our model needs to be able to encode them so that both representations are close in the vector space of their representations (Figure~\ref{fig:proposedmodel}). To further enforce closeness between both the client and the corresponding standardized representation, we leverage a contrastive learning mechanism \cite{Hadsell_2006} that uses an adaptation of the N-Pair loss function\cite{Sohn_2016}. 
In the retrieval/matching phase, our approach helps, given an initial client description,  to rank the respective SD and associated DP  among the first positions.\looseness=-1

\vspace{-0.45cm}
\vspace{-0.45cm}

For doing so, we propose a novel training strategy at training time in which the transformer encoders train cooperatively, optimizing their parameters at alternating training steps mirroring a two-player turn-based game. We call this mechanism \textbf{TaG-Training} (\underline{T}urn-based B\underline{a}rgaining \underline{G}ame \underline{T}raining), as detailed in Section \ref{tag_training}. Our results revealed that tag training ensures a more accurate and fine-grained optimization. Efficiency is also improved from the optimization perspective by leveraging a learning-based loss function~\cite{Bellet_2015,Huai_2019,Mahmut_2019}: our encoders aim to learn nonlinear embedding representations of the input languages that place similar data points close while positioning dissimilar ones far on the embedding space~\cite{Himansu_2020, Sohn_2016}. Previous work has exploited either contrastive loss~\cite{Hadsell_2006} or triplet loss~\cite{Hoffer_2015}  that suffer from slow convergence, partially because they exploit only one negative example at each update step. Differently, within \textbf{TPDR}, we propose to take advantage of an adapted \textbf{N-pair loss} function, which generalizes triplet loss by allowing joint comparisons among more than one negative example. For ranking tasks such as product retrieval, $N$-pair loss is more effective than other ranking losses, as demonstrated in our experiments because a single update step pulls up relevant products for its query while pushing down $(N-1)$ non-relevant products. The number of non-relevant products for a given query usually greatly surpasses the relevant ones, making a significant difference.\looseness=-1

After generating this initial ranking, we proposed an additional step: re-ranking. Unlike traditional two-phase (ranking and re-ranking) IR approaches, where an initial ranking is first generated from a "weak" approach and, then, is applied a re-ranking based on a neural approach~\cite{two-step-example}, our approach performs the inverse process. Our approach initially generates the initial ranking based on a strong neural network process (described above). This first step reduces the number of candidates considerably. After that,  we apply  a re-ranking step based on syntactic features that are very important for the exact matching (model, dimension, size) of certain products but may have been neglected by the transformers. In this way, our solution innovates by exploiting both aspects: semantics and syntax.  We propose to consider as the first metric the similarity value of the contextual model (without {\it re-ranking} step), that is, the weighting of the output of the previous prediction phase. Note that, given the various advantages arising from the use of the proposed contextual model, in order to enable improvements in terms of effectiveness, it is essential to reuse the weighting of the pre-stage model of {\it re-ranking}. Furthermore, we consider three additional syntactic metrics: (1) TFIDF Cosine Similarity; (2) Jaccard Textual \textit{bigrams} Comparison; and (3) to further improve the results, the BM25 score.\looseness=-1

To evaluate our proposal, we consider three real domains -- consisting of 11 datasets -- from the Astrein company, covering different application contexts and dataset sizes. As a result, considering a \textbf{general application context} covered by the first nine distinct datasets (Section~\ref{results_general}), with a product catalog (search index) consisting of almost 500 thousand products, the developed solution was able to retrieve the actual final product description before the $5^{th}$ ranking position in \textbf{71\%} of the cases (on average). Furthermore, we observe that in \textbf{85\%} of the cases, the respective standardized product is retrieved before the $100^{th}$ position of the rank. In cases where we could not retrieve the standardized product in the first positions of the ranking, the correct association of DPs is of paramount importance since this process facilitates the manual search of the real item constituted only by products of the same DP. In fact, the proposed solution can retrieve the correct DP in the first ranking position for \textbf{80\%} of queries (in a search universe consisting of \textbf{3972} distinct DPs). Furthermore, the correct associated DP is contained up to the $5^{th}$ ranking position in \textbf{88\%} of queries, representing an excellent practical result in the presented domain.\looseness=-1 

Finally, in the other domains (D10 and D11 datasets), both consisting of products associated with a single DP, namely \textbf{Orthoses, Prostheses and Medical Materials - OPMM} and \textbf{Medicines}, the results confirm the effectiveness of the proposed model and demonstrate its flexibility. More than 70\% of standardized products are present up to the $5^{th}$ ranking position for both scenarios. We also evaluate each step of our proposal individually, demonstrating the complementary between syntactic and semantic information to properly deal with the product standardization problem.\looseness=-1 
\section{Background}
\label{sec:background}

\subsection{Self-Attention}\label{sec_self-att}

Self-attention is a mechanism that allows finding meaningful words in a sentence given a query word in the same sentence (hence the "self"). The objective is to capture those words that contribute the most to the semantics of the query word. Mathematically, self-attention is simply a function that receives a sequence $X$ of $N$ vectors $x_i \in R^d$ as input and returns as output another sequence $Y$ of $N$ vectors $y_i \in R^d$, where each vector in $Y$ is simply a weighted average (the weights are called attention scores) of the vectors in $X$.\looseness=-1

\vspace{-0.3cm}
\begin{figure}[htb]
\centering
\includegraphics[width=0.4\textwidth]{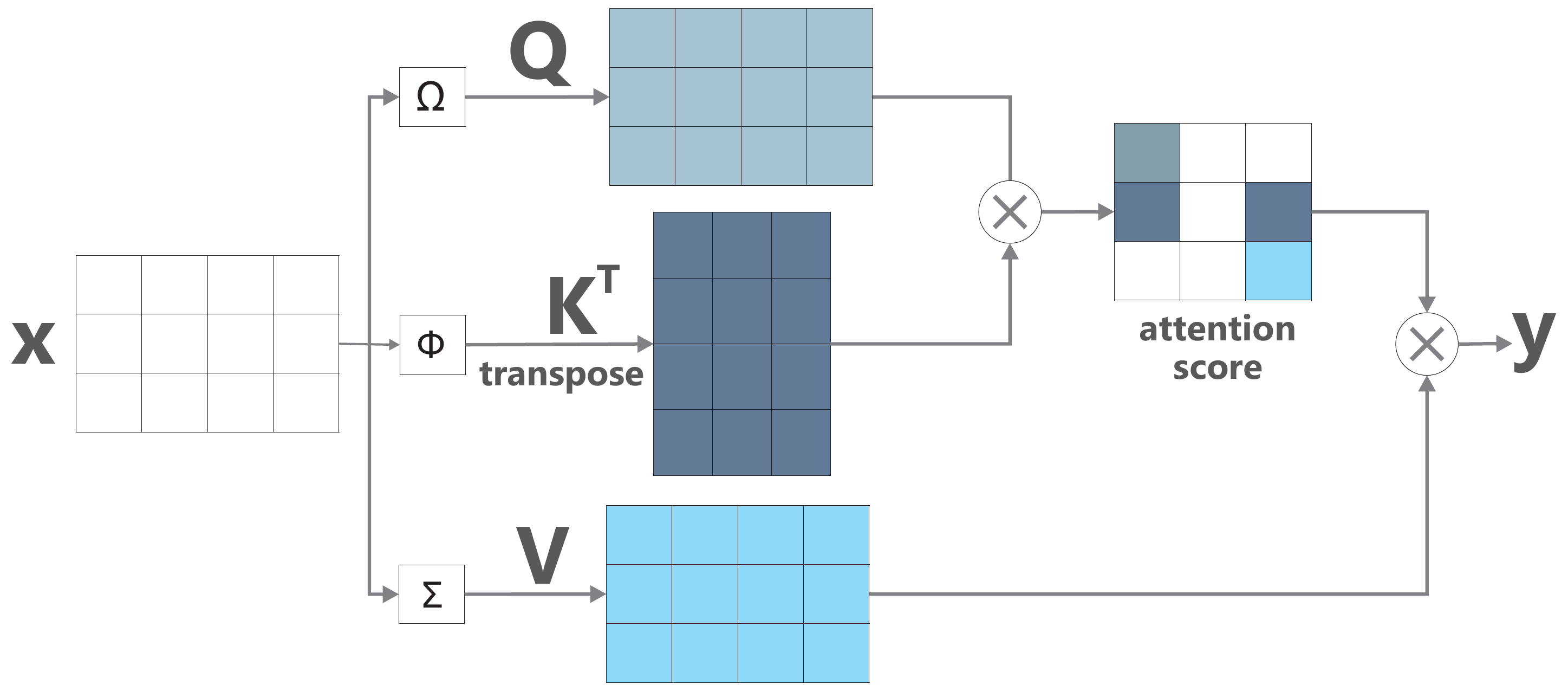}
\vspace{-0.3cm}
\caption{\footnotesize{Architecture of the self-attention deriving key, query, and value representations from input X, calculating attention scores and outputting the final representation Y.}}
\label{fg_self_attention}
\end{figure}
\vspace{-0.3cm}

As shown in Figure~\ref{fg_self_attention}, from the input sequence $X$, distinct feed-forward neural networks output Q (Query), K (Key), and $V$ (Value) matrices, and the attention scores are calculated as defined in Equation~\ref{attention_scores}. Finally, the output $Y$ is obtained by multiplying the attention scores by matrix $V$\looseness=-1

\vspace{-0.3cm}
\begin{equation}\label{attention_scores}
    Attention(Q,K,V)=softmax \left( \frac{QK^T}{\sqrt{d_k}} \right) V
\end{equation}

\noindent where the scaling factor $d_k$ is the dimension of vectors in $K$.\looseness=-1

\subsection{Multi-Head Self-Attention}\label{sec_multi_head_self_att}

Multi-head self-attention involves applying different attention mechanisms to the same input sequence $X$. Each head can extract different characteristics from the input sequence. In Figure
~\ref{fig:proposedmodel}, for instance, when using multi-head self-attention to represent the fine-grained \textit{retainer} specification, while one head focuses on the \textit{retainer} word, another head can focus on the words defined by product specification, such as length (\textit{83.73x114.3x11.9mm}) and material (\textit{70sha}). The model can capture multiple patterns over both product specifications by leveraging multi-head self-attention. The independent self-attention outputs are then concatenated and linearly transformed to generate the $Y$ output.\looseness=-1

\subsection{Transformer's Encoder}\label{sec_transformers_encoder}

Proposed by Vaswani et al.~\cite{Vaswani_2017}, the transformer emerges as an answer to the CNNs and RNNs drawbacks. Specifically, while RNNs do not allow parallelization during training because of their sequential nature and suffer from vanishing gradients, when using CNNs it is challenging to find relevant patterns in distant positions within the input sequence \cite{Vaswani_2017}.\looseness=-1

Transformer dispenses recurrence and convolutions entirely and improves the sequence representation by adopting only the self-attention mechanism. Bert~\cite{Devlin_2019}, GPT~\cite{BBrown_2020}, \textit{RoBERTa}~\cite{Liu_2019}, and other transformer-based models have been the state-of-the-art for a myriad of Natural Language Understanding tasks such as question-answering, named entity recognition, text classification, and other~\cite{Cunha_2021,GargVu_2020,Cunha2023A}.\looseness=-1

\begin{figure}[htb]
\centering
\includegraphics[width=0.20\textwidth]{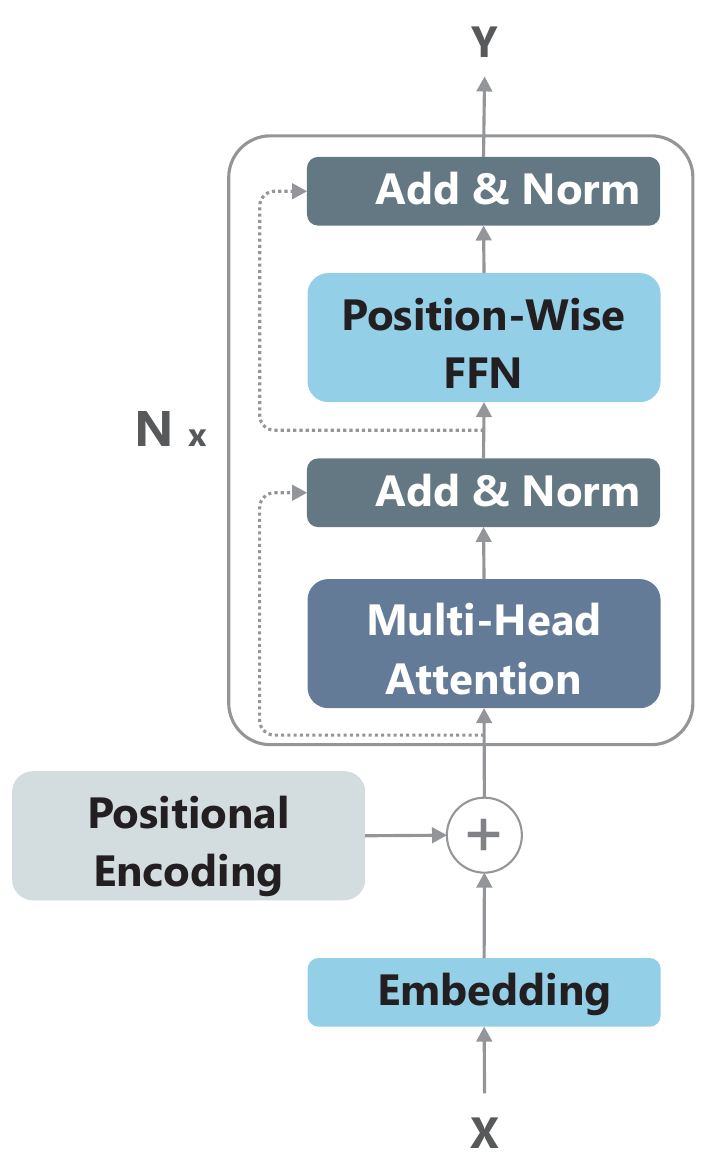}
\vspace{-0.5cm}
\caption{\footnotesize{The transformer's encoder architecture.}}
\label{transformer_architecture}
\end{figure}

The transformer follows the architecture of $N$ identical blocks composed of two layers: a multi-head self-attention mechanism and a simple position-wise fully connected feed-forward network, as shown in Figure~\ref{transformer_architecture}. There is also a residual connection~\cite{He_2016} around each layer, followed by a normalization layer.\looseness=-1

\vspace{-0.5cm}
\vspace{-0.3cm}

Before forwarding through the $N$ blocks, each sequence token is passed through a learned embedding layer. Next, as the meaning of a word depends on its place in the sentence, its embedding should encode its position. To this end, the positional encoding layer adds numbers between $\left[-1,1\right]$ using non-learned sine and cosine functions to the embedding tokens.\looseness=-1
\section{The Proposed Approach}
\label{sec_model}

\begin{figure}[htb]
\centering
\includegraphics[width=0.37\textwidth]{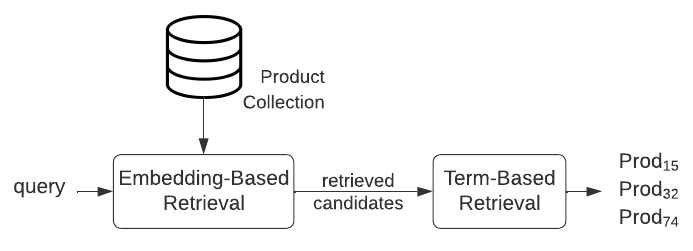}
\caption{The overall architecture of \textit{TPDR}.}
\vspace{-0.3cm}
\label{proposed_d_architecture}
\end{figure}

Figure \ref{proposed_d_architecture} shows the \textbf{TPDR} architecture in which, given a  query (i.e., initial client description), we first use an embedding-based retrieval system to retrieve a reduced list (when compared to the whole catalog) of  potentially relevant standardized products based on semantic matching. As we observed that in practice, most relevant products could not be retrieved at the top of the ranking due to syntactic mismatches, we use a term-based retrieval to re-rank the (query product) candidate pairs in the second stage to output the final product ranking that better matches specific characteristics of the desired product (e.g., model, dimensions).\looseness=-1

\subsection{Embedding-based Retrieval}
\label{sc_code_search_engine}

\begin{figure}[htb]
\centering
\includegraphics[width=0.28\textwidth]{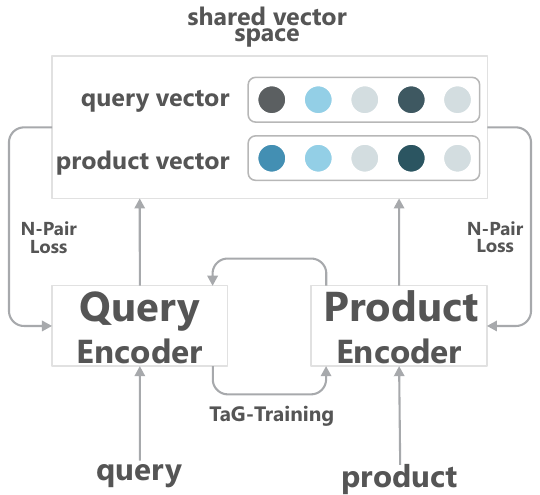}
\vspace{-0.3cm}
\caption{The overall architecture of \textit{TPDR}.}
\label{proposed_architecture}
\end{figure}

We employ dual transformer encoders (Section 2.3) for our Embedding-based semantic Retrieval. These encoders work as a logical framework to represent the queries, and product descriptions in the same embedding space as Figure~\ref{proposed_architecture} illustrate. At indexing time, our search engine indexes the product catalog description collection encoded by the \textit{Product Encoder}. At the searching time, the \textit{Query Encoder} should only represent a single query. Then, the closest $k$ standardized products concerning the query are selected from the collection using a similarity search method.\looseness=-1

We also optimize both encoders alternately,  \textit{Query Encoder}'s parameters in one training step and the \textit{Product Encoder}'s parameters in the next step. We call this mechanism \textbf{TaG-Training} (as explained in Section \ref{sc_tag_training}). It resembles a two-player turn-based game where the combined action of the players contributes to optimizing the payoff (or minimizing the loss function)~\cite{Maschler_2020}. Conveniently, the training exploits the $N$-pair loss function that is suitable for the product retrieval task, as explained in Section \ref{subsec:npair}.\looseness=-1

\vspace{-0.3cm}
\vspace{-0.3cm}

Each \textit{TPDR} encoder is a transformer with $12$-layers, generates a $768$-dimensional vector representation for each token from the input language, uses a $12$-head self-attention, and contains $110$M parameters. As input, each encoder receives a token sequence resulting from the byte-pair encoding (\textit{BPE}) \cite{Gage_1994, Sennrich_2016} of the respective input languages. The transformer returns a sequence of hidden states from the model's last layer as output. Finally, an average pooling over the last hidden states generates a single $768$-dimensional dense vector representing the input language.\looseness=-1

To complete our end-to-end product search engine, the fine-tuned \textit{Product Encoder} is employed to generate product embeddings from the collection. These embeddings are indexed by leveraging Hierarchical Navigable Small World graphs
~\cite{Malkov_2020} that offer a  better logarithmic complexity scaling and allow an efficient nearest neighbor search. Finally, a query is answered by generating a query embedding (using \textit{Query Encoder}) and returning its nearest $k$ products.\looseness=-1

\subsection{Term-Based Retrieval}

Our main motivation for proposing a re-ranking strategy was that in our preliminary experiments, although $85\%$ of the final product descriptions were present up to the $100^{th}$ ranking position, in several  cases, relevant products were not retrieved in the top positions of the ranking. This is due to the existence of several similar products in the catalog and a ´´syntactic mismatch´´ between the client description and relevant standardized description regarding specific characteristics such as the model or dimensions of the product that is better captured by a syntactic match.
Thus, from the  list of candidate products received from the Embedding-based retrieval, we use a Term-based retrieval to re-rank the product query pairs and generate the final product ranking in the second step, aiming to solve the aforementioned issues.

In more detail, our Term-based retrieval consists of a linear combination of pre-fixed scores to improve the retrieval results. Specifically, we considered four ranking functions: the Embedding-base Retrieval ($s_1$) score (from the previous step), cosine similarity ($s_2$), Jaccard similarity ($s_3$), and BM25 ($s_4$) score.

\paragraph*{\textbf{Cosine Similarity}~\cite{simcos}:} we first leverage bag-of-words representation based on TF-IDF weighting-scheme for representing query and product as a sparse vector, and then we measure the cosine similarity of them using the Equation \ref{simcos}.

\begin{equation}\label{simcos}
    s_{2} (\vv{q} {,} \vv{p}) = \frac{\vv{q} . \vv{p}}{\mid \mid \vv{q} \mid \mid . \mid \mid \vv{p} \mid \mid }
\end{equation}

\paragraph*{\textbf{Jaccard Similarity}:}  we first convert the query and product description into two sets of bigrams, then measure the similarity between two sets: capturing which members are shared and which are distinct. Finally, we calculate the Jaccard similarity by dividing the size of the intersection divided by the size of the union of two sets, as shown in Equation~\ref{jacc}.

\begin{equation}~\label{jacc}
   s_3 (Q, P) = \frac{Q \cap P}{Q \cup P} 
\end{equation}

\paragraph*{\textbf{BM25}~\cite{bm25}($s_4$):} similar to the cosine similarity, we leverage the TF-IDF to measure BM25 (Eq. \ref{eq_bm25}) to score each pair query product.

\paragraph*{\textbf{Final Weighting}}

The final ranking score $S$, (Equation \ref{f_ranking}), corresponds to the linear combination of the previous ranking functions.

\begin{equation}\label{f_ranking}
  S(Q,D) = \frac{[3 * s_1(Q,P) + s_2(Q,D) + s_3(Q,D) + s_4(Q,D)]}{6}  
\end{equation}

Note that, as the results of the contextual Embedding model were already considerably effective, we weighted them with larger importance in the final model (3 times higher than the syntactic ones). We leave for future work to adjust the weights of each component for each dataset. 
\looseness=-1

\subsection{Learning Objective}
\label{subsec:npair}

A contrastive learning method is particularly suitable for IR tasks because it pulls closer entities with similar semantics, such as a developer's query and its relevant product, and pushes away others with distinct semantics. To optimize the encoders' parameters, we leverage an adaptation of $N$-pair~\cite{Sohn_2016}, a loss function that demonstrates superiority for a variety of IR tasks, including fine-grained object recognition~\cite{Xie_2015}, image clustering~\cite{Song_2016}, and face identification~\cite{Schroff_2015}.\looseness=-1

More formally, consider $N$ training samples $\{(q_1, c_1), ... (q_N, c_N)\}$, where $q_i$ is a query in natural language perfectly answered by the product $c_i$. The core contribution of the $N$-pair loss described in Equation~\ref{n_pairs_loss}, is to optimize the embedding functions $f_i=f(q_i;\theta)$ and $g_i=g(c_i;\sigma)$ to be able to map $q_i$ and the corresponding $c_i$ into dense vectors where the similarity $sim(f_i, g_i)$ is greater than the similarity $sim(f_i, g_j)$; $g_j$ being the embedding of any other product not answering $q_i$. That is, at the same step that a relevant product is placed closer to a query, it is also pushed further away from \textit{(N-1)} non-relevant products.\looseness=-1 

\begin{equation}\label{n_pairs_loss}
    \mathcal{L}(q_i,c_i, \{c_j\}) = -log \frac{exp(f_i^T g_i)}{exp(f_i^T g_i) + \sum_{j=1}^{N-1}exp(f_i^T g_j)}
\end{equation}
\noindent where $c_i$ is a relevant product with respect to $q_i$ and all other products $c_j$ (with $i \neq j$) are non-relevant ones.\looseness=-1 

\begin{figure}[h!]
\centering
\includegraphics[width=0.38\textwidth]{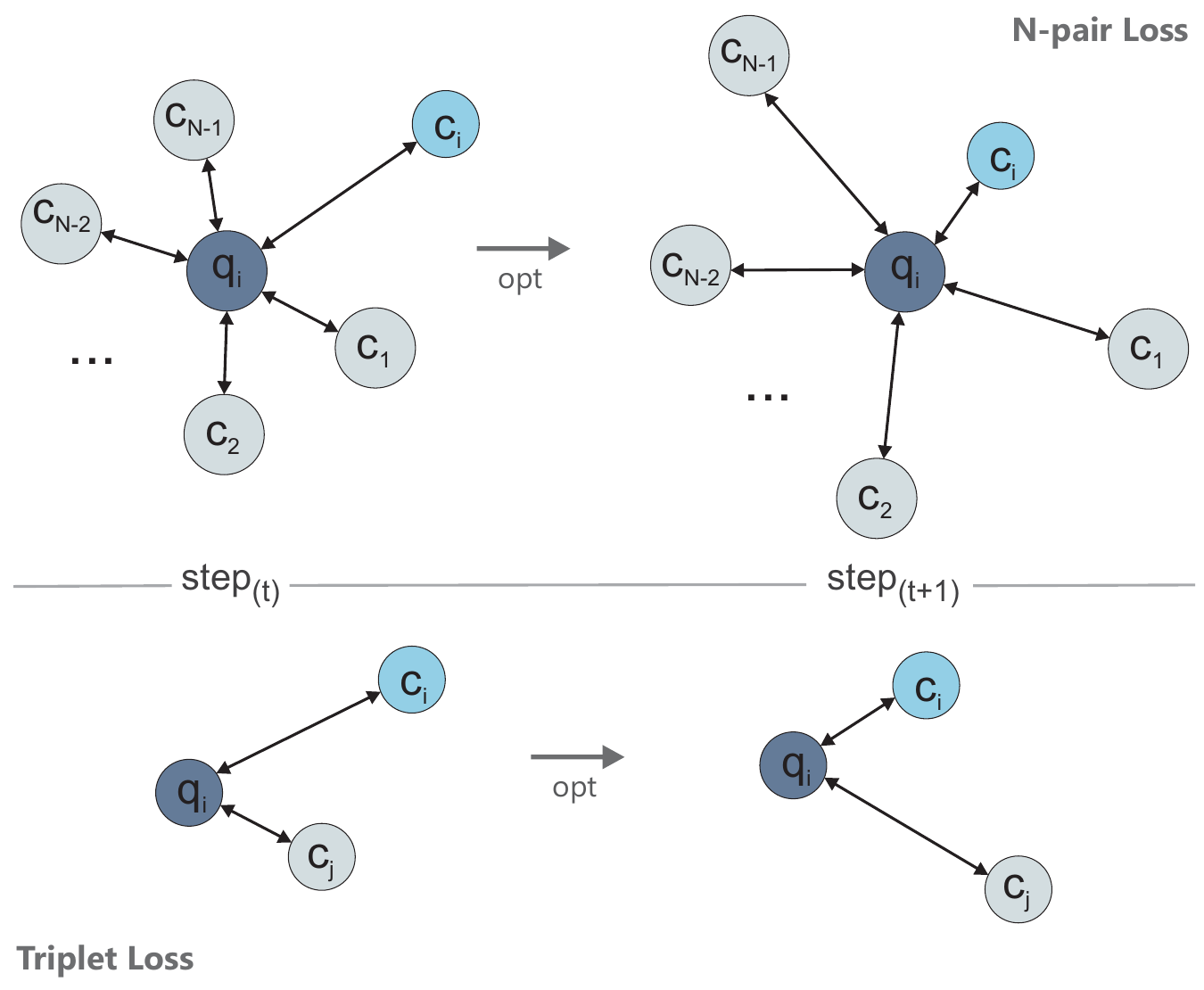}
\caption{Embeddings of a model trained to satisfy the constraints of \textit{triplet} loss (bottom) and $N$-pair loss (top).}
\label{n_pair_loss}
\end{figure}

Other ranking losses such as \textit{contrastive} loss~\cite{Hadsell_2006} and \textit{triplet} loss~\cite{Hoffer_2015}, widely used in product retrieval, employ only one negative example, not interacting with the other negative samples in each update step, as shown in Figure~\ref{n_pair_loss}. Vertices are representations of queries and products, and the edges express the distance. The triplet loss pulls the positive example while pushing one negative example at a time. On the other hand, the $N$-pair loss pushes $(N-1)$ negative examples all at once. Reaching the same effect would require an inefficient batch construction strategy resulting in $N(N-1)$ pairs when using triplet loss.\looseness=-1

$N$-pair loss is very suitable for the product retrieval task, characterized by a few (or a single) relevant items in a collection of numerous irrelevant ones. It also converges faster, uses an efficient batch construction strategy, and the use of multiple negative examples enables the encoders to generate discriminative representations.\looseness=-1 

\subsection{TaG-Training}
\label{sc_tag_training}

Unlike in Generative Adversarial Networks (\textit{GANs})~\cite{Ian_2014}, where the training corresponds to a minimax two-player game, the training of the \textit{Product} and \textit{Query Encoders} mirrors a two-player turn-based bargaining game in which the joint actions players take result in collective payoffs.\looseness=-1 

Specifically, the optimization of the \textit{Query Encoder} causes it to move the noisier natural language's representation closer to the standardized natural language representation. In the same optimization step, the \textit{Product Encoder} moves the product representation closer to the region in the space where the query representation lies before optimization. Then, updating both encoders at the same training step may not be the best solution as the model's parameters may oscillate and have sub-optimal convergence.\looseness=-1

Employing an optimizer for each encoder promotes a more refined parameter update. We propose that updates occur in turns, with one encoder per optimization step. Therefore, each encoder behaves like a player, playing alternate turns to minimize the loss function. We call these mechanisms \textbf{TaG-Training} (\underline{T}urn-based B\underline{a}rgaining \underline{G}ame \underline{T}raining).\looseness=-1 

\begin{figure}[htb]
\centering
\includegraphics[width=0.35\textwidth]{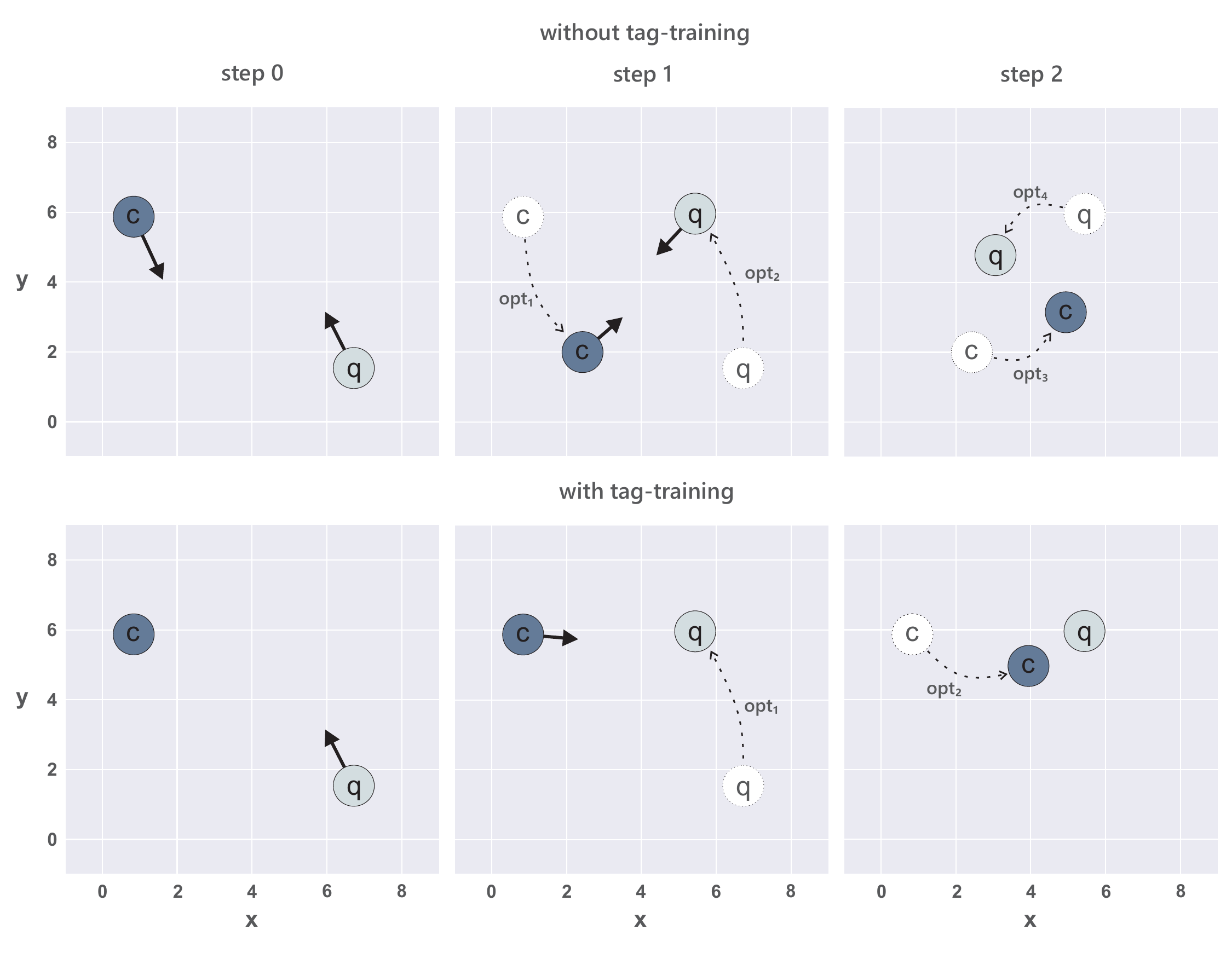}
\vspace{-0.3cm}
\caption{Three successive optimizing steps of the encoder’s parameters without and with tag-training.}
\label{tag_training}
\end{figure}

Figure~\ref{tag_training} shows an illustrative example of three successive optimizing steps of the two encoders' parameters without and with tag-training. Both approaches have the same initial configuration in which we have a product embedding $c$ and a query embedding $q$ in a two-dimensional space. The loss emphasizes the distance between these two embeddings, and the arrow represents the gradient.\looseness=-1

\vspace{-0.3cm}
\vspace{-0.3cm}

Without tag-training, the optimizer updates both encoders' parameters at all training steps. In step $1$, the \textit{Product Encoder's}
optimization (opt1) drives $c$ to the surroundings of $q$. A position miss-match occurs because, in the same step, the \textit{Query Encoder's} optimization (opt2) moves $q$ to be closer to a position where $c$ was. The same problem occurs in step $2$ and in the following steps, where $c$ and $q$ orbit each other before converging.\looseness=-1

With tag-training, optimizing the encoder's parameters occurs in alternating steps. The initial position of $c$ and $q$ in step $0$ is the same as in the previous scenario. The difference occurs in step $1$ where only the \textit{Query Encoder} is optimized (opt1), moving $q$ to a position closer to $c$, which remains unchanged. In the last step, the \textit{Product Encoder's} parameters are optimized (opt2), placing $c$ nearby to $q$.\looseness=-1

TaG-training can potentially improve effectiveness, yet its direct impact is on efficiency. Leveraging tag-training not only allows a fine-grained joint optimization of each encoder but is also responsible for reducing the amount of the encoder's parameter updates and, ultimately, the overall training time. Alternating between both encoders in the optimization step, \textit{TPDR} has an equal training time if it had used a single transformer. 
Regarding effectiveness, the joint optimization with fewer parameter updates may help avoid overfitting, contributing to the model's generalization.\looseness=-1

\subsection{TaG-training as a Two-player Bargaining Game}\label{appendix_a}

Let $G=(S, d)$ be a two-player bargaining game   where~\cite{Maschler_2020}:\looseness=-1

\begin{itemize}
    \item $S \subseteq {\rm I\!R}^2$ is a nonempty, compact, and convex set called the set of alternatives.
    \item $d = (d_1,d_2) \in S$ is called the disagreement point (or conflict point).
    \item There exists an alternative $x = (x_1, x_2) \in S$ satisfying $x > d$ (e.g., $x_1 > d_1$ and $x_2 > d_2$)
\end{itemize}

We interpret the tag-training as a bargaining game where players \textit{Query Encoder} and \textit{Product Encoder}
need to agree on an alternative $x = (x_1, x_2) \in S$. If they come to such an agreement, \textit{Query Encoder's} payoff is $x_1$, and \textit{Product Encoder's} payoff is $x_2$. If the players cannot agree, the game's outcome is $d=(0,0)$ (the encoders' payoff is $d_1$ and $d_2$, respectively).\looseness=-1

The utility function for the \textit{Query Encoder} (and the \textit{Product Encoder}, by symmetry) defined in Equation \ref{utility_function} measures the proportion of query $q_i$ in which the relevant product $c_i$ could be found up to the $k$ ranking position. By taking $k=1$ we are interested in the proportion of queries where $c_i$ is closest to $q_i$ for all $i \in \{1,2, ...N\}$. Therefore, the set of possible agreements is $S=\{(x,x): 0 \leq x \leq 1\}$, and the vector of disagreement is $d=(0, 0)$.\looseness=-1 

\begin{equation}\label{utility_function}
    Recall@k = \frac{1}{|N|}\sum_{i=1}^{|N|} \sigma(p_i)
\end{equation}

\noindent where $p_i$ is the position $c_i$ in the ranking of the closest points relative to a query $q_i$ and $\sigma$ function returns $1$ if $p_i \leq k$, otherwise it returns $0$.\looseness=-1

Equation~\ref{nash_solution} defines the point $N(S, d)$ called the Nash agreement point (or the Nash solution, first proposed by John Nash in 1953)~\cite{Nash_1953}. For $G$,  the alternative $x = (1,1) \in S$ is the Nash solution.\looseness=-1 

\begin{equation}\label{nash_solution}
    argmax_{x \in S, x \geq d} (x_1 - d_1)(x_2 - d_2)
\end{equation}

From an optimization perspective, the global minimum $M^*$ of the $N$-pair loss occurs when $f_i^Tg_j$ becomes infinitely small ($f_i^Tg_j \to -\infty$): the distance between a query and any non-relevant product tends to infinity. Although this theoretical result cannot be achieved in practice, the purpose of $N$-pair loss is to approximate a query $q_i$ to its relevant product $c_i$ while pushing away the non-relevant products $c_j$ for all $i,j \in \{1,2,...,N \}$ and $i \neq j$.\looseness=-1 

Therefore, any configuration where the query $q_i$ is closer to $c_i$ than any other points $c_j$ constitutes a solution $M$ to the product retrieval task that can be achieved via gradient descent. Since $M$ causes $x = (1,1) \in S$, it is a Nash agreement point for $G$.\looseness=-1

\section{Experiments}\label{sec:res}

In this section, we will present the results obtained from the instantiation of our proposal. All experiments were performed using a machine with the following hardware configuration: 64GB of RAM, 16 vCPUs, and 1 NVIDIA 3090 GPU with 24GB.\looseness=-1

\subsection{Data}

As previously mentioned, the proposed model receives a set of pairs as input (query, response). In our domain, we represent pairs (\textbf{Initial Customer Specification -- Query, Standardized Description - Product}). In addition, each product is associated with a class named \textit{Description Pattern - DP}, which is equivalent to a class or category of products. As mentioned, the correct association of this DP is important because, in cases where we cannot retrieve the real products in the first positions of the ranking, this process facilitates the manual search of the real item consisting only of products from the same DP. Next, we describe the data inputs generated for the training of the proposed algorithm.\looseness=-1

The first set of datasets (D1--D9) -- general purposes -- consists of matching items (query and product) of different dimensions\footnote{The amount of data in each dataset is described in parentheses. For instance, D1 is composed of 20,707 (query, product) pairs.} referring to diverse companies. This dataset is associated with \textbf{3,972} distinct DPs. This set of datasets is associated with a shared search index (catalog) composed of \textbf{441,223} different products.\looseness=-1

The \textbf{D10} dataset consists of 21,225 pairs referring to a single DP - \textbf{OPME - Ortheses, Prostheses, and Medical Materials} and has a search index of 1 million different products. Thus, even though being associated with only one DP domain, this dataset is constituted of a few training pairs compared with the product catalog size. Indeed, the search space is considerably larger than the previous case, making it a more challenging scenario. Finally, the \textbf{D11} dataset consists of 2 thousand pairs (query and product) referring to a single DP - \textbf{Medicines}. As a search index, we adopted a product catalog containing approximately 29,000 different products.\looseness=-1

Following the scientific methodology present in \cite{Codexglue_2021,Leclair_2019}, we randomly divided the dataset pairs into 80\% for the training phase; 10\% for the validation phase (used for the calibration process and parameter refinement); and 10\% for the test phase.\looseness=-1

\subsection{Evaluation Metrics}

In our results analysis, we considered three traditional retrieve metrics: (i) \textit{Mean Reciprocal Rank} (MRR)~\cite{Aslanyan_2020} required to assess in what ranking position the relevant products tend to be placed and want that item to be at a higher position (MRR is very informative since we have one relevant product per query); w(ii) \textit{Normalized Discounted Cumulative Gain} (nDCG)~\cite{Karmaker_2017} which penalize highly relevant products when appearing lower in ranking; and(ii) Recall~\cite{Li_2022} to measures the proportion of relevant products retrieved.



Therefore, given a pair \textit{(Initial Customer Description, Standardized Description)}, \textit{Initial Customer Description} corresponds to a query in which we consider the \textit{Standardized Description} as the only relevant result -- a challenging scenario compared to traditional search situations, as we only have a single relevant result per query.\looseness=-1

\subsection{Proposed Model Instantiation}

As mentioned, the proposed model is based on a BERT architecture for encoding descriptions via both encoders (query and product). Considering the validation set, we perform a preliminary set of model refinement experiments by varying the set of the model´s hyperparameters. The best variation of hyperparameters according to these preliminary experimentations of the proposed model is described below:\looseness=-1

\begin{itemize}
     \item Max number of training epochs: 10
     \item Client Initial Description max length: 64 terms
     \item Standardized Product Description Max length: 64 terms
     \item Batch size: 32
     \item Pre-trained Embedding: \textit{sentence-transformers/LaBSE} 
     \item Re-Ranking: i. Prediction Phase Score; ii. Cosine similarity; iii. Jaccard Comparison of bigrams; iv. BM25.
\end{itemize}

Given that both descriptions (initial and standardized) are mostly written in Portuguese and English, we propose the use of \textbf{LaBSE}~\cite{feng2020language} \textit{embeddings} for two main reasons: i. the results of pre-training with this embedding model were considerably higher than those obtained with  alternatives such as \textit{BERTimbau}~\cite{souza2020bertimbau} and \textit{MUSE}~\cite{yang-etal-2020-multilingual}; and ii. this particular type of \textit{embedding} can capture term relationships at a multi-lingual level. As we will show below, the use of this model can provide higher effectiveness and deal with cases of the same word  written in different languages (e.g., \textit{paper} and \textit{papel} in Portuguese).\looseness=-1

Finally, at the description indexing phase, we opted for using the exact solution (\textit{brute\_force} method) in order to optimize the generation of indexes generated by the HNSW - \textit{nmslib}~\cite{Malkov_2020} library. The use of this method provides considerable improvements in terms of effectiveness (MRR). Preliminary results varying the set of hyperparameters showed that \textit{brute\_force} is up to 89\% more effective than the approximate method \textit{hnsw} in the considered product retrieval context. However, when effectiveness is prioritized, efficiency may be penalized. Despite that, considering that the validation set had 10K queries, the \textit{brute\_force} method represented an increase of only 15 minutes in the description retrieval step. Therefore, we believe using the exact solution provides a better tradeoff between effectiveness and efficiency.\looseness=-1

\subsection{Baselines}



\subsubsection{Classical Syntactic Search Engine}

As a syntactic baseline, we rely on a traditional probabilistic model, which 
estimates the probability that a product will be relevant to a 
query. We apply $TF\text{-}IDF$ as a weighting scheme, a logical framework for representing documents and queries, and BM25 as a ranking function. 
\textit{BM25} is a family of ranking functions that scores a set of products based on relative proximity between the query within product description terms. 
Eq. \ref{eq_bm25} shows one of the most prominent instantiations of the \textit{BM25} given a query q containing keywords $q_1, ..., q_n$ and a product $p$.\looseness=-1

\vspace{-0.3cm}
\begin{equation}\label{eq_bm25}
sim_{BM25}(q,p)=\sum_{i=1}^{n}TF\text{-}IDF(q_i).\frac{f(q_i,p)(k_1+1)}{f(q_i,p)+k_1(1-b+b\frac{|D|}{\overline{d}})}
\end{equation}

\noindent where $f(q_i,p)$ is the $q_i$ frequency in d,  $\frac{|D|}{\overline{d}}$ is the relative product description length over the average product description length in the collection. $k_1$ and $b$ are free parameters, usually chosen, in the absence of optimization, $1$ and $0.75$, respectively. Finally, $TF\text{-}IDF(q_i)$ is the $TF\text{-}IDF$ weight of the query term $q_i$. 

\subsubsection{Pure Semantic Search Engine} 
Classical search engines treat product descriptions in the same way traditional search engines deal with natural language documents. In the face of the \textit{vocabulary mismatch}~\cite{Gysel2018}, they are unable to match the user's needs with the relevant products. Recent studies have exploited models that learn to encode both inputs into a shared vector space. Ideally, relevant product vectors will be closer to the respective query vectors (while the irrelevant ones will be farther away). 
Therefore we defined the \textbf{TPDR (noRR)} -- our proposal without the re-ranking step --, fine-tuning a LaBSE transformer encoder to represent queries and products with the same vector space, and we adopted the same parameters used in our approach. TPDR (noRR) also generates product embeddings from the product collection and indexes them using HNSW library. Finally, a query is answered by generating by encoding a query and returning its nearest $k$ products.



\subsection{Experimental Results}

\subsubsection{\textbf{Effectiveness Analysis of the client description match with a  previously standardized item in a product catalog}}~\label{results_general}

In Table \ref{tab:effec_results}, we present the effectiveness results of our approach using the MRR, NDCG, and Recall metrics. A first important observation is that the purely semantic approach, which corresponds to the first step of our method -- \textbf{TPDR (noRR)}) -- in almost all cases, considering all datasets and metrics, with a few exceptions (e.g., D7 and D9) is better, sometimes by large margins (up to 6x as in D2), than the purely syntactic approach (BM25), demonstrating the importance of our innovations regarding the semantic aspects of the problem.

\vspace{-0.7cm}
\begin{table}[!h]
    \centering
        \begin{figure}[H]
        	\centering

        	\includegraphics[width=0.42\textwidth]{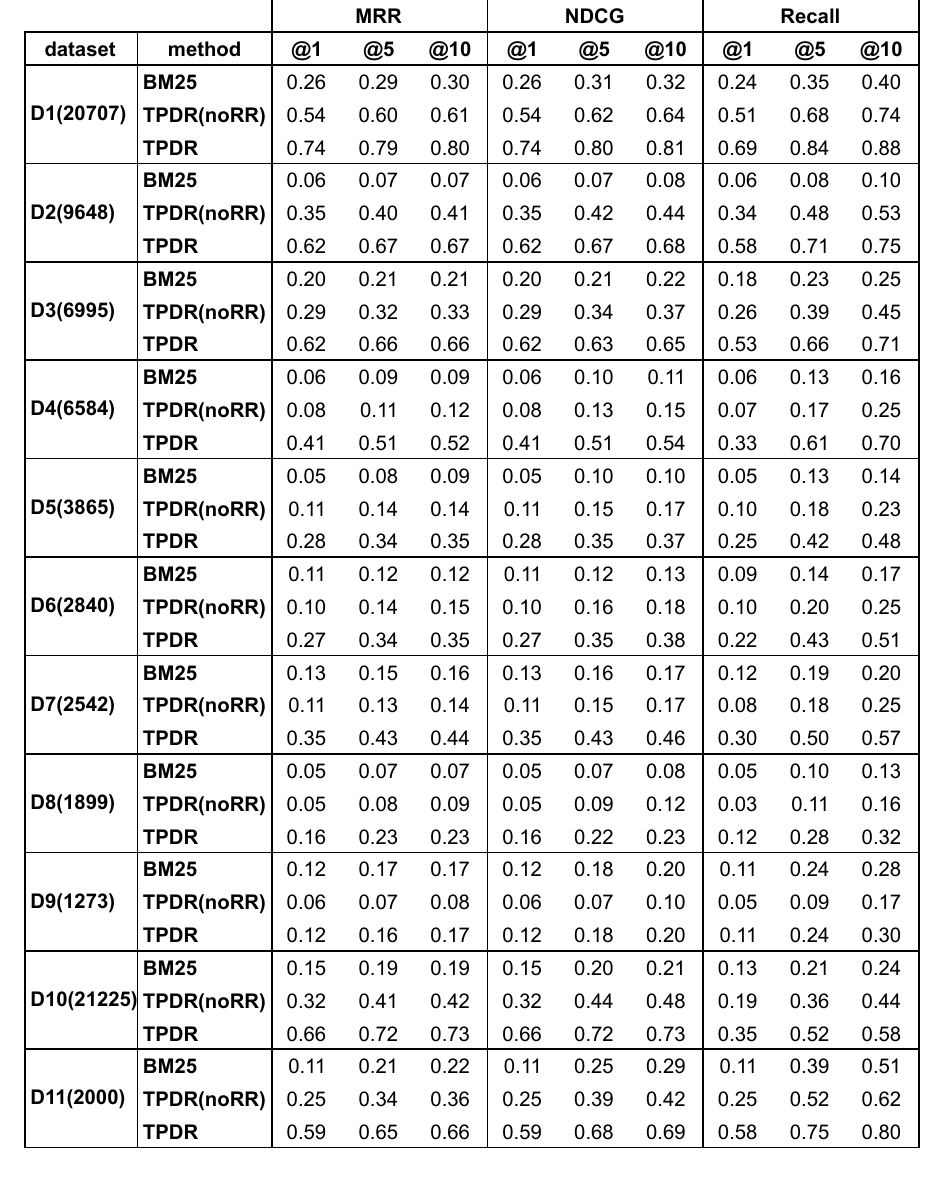}
        \end{figure}
    \vspace{-0.5cm}
    \caption{Effectiveness -- MRR, NDCG and Recall metrics.}
   \vspace{-0.7cm}
    \label{tab:effec_results}
\end{table}

Regarding our complete proposal \textbf{TPDR}, MRR@1, for instance, shows that considering the mean of the queries of all datasets, 64\% of the standardized descriptions present in the test datasets are retrieved, on average, in the first position -- varying between 11\% in (D9), which has the lowest number of samples and 74\% in (D1), the largest dataset of the first scenario (several DPs). This behavior is evidence that the number of pairs for training has a direct impact on the effectiveness of the proposed model. Furthermore, MRR@5 and MRR@10 results show that the actual standardized descriptions retrieved by our approach are present, on average, between the first and second positions for all queries\footnote{An average MRR value of 0.5 is produced as on average the actual result appears in the second position, i.e., $\frac{1}{rank = 2}$}. Finally, still considering the MRR metric (in variations $@1$, $@5$ and $@10$), our proposed model was able to produce average improvements of \textbf{3.76} times -- ranging from \textbf{1.01} times to \textbf{9.8} times -- when compared to the pure syntactic approach (BM25). And when compared to the pure semantic approach (\textbf{TPDR (noRR)}), the complete \textbf{TPDR}  instantiation  produced  average improvements of \textbf{2.3} times, demonstrating the importance of the syntactic-based re-ranking.\looseness=-1

\textbf{TPDR} is also very effective when considering  \textit{nDCG} and \textit{Recall}. In the most challenging scenario (the first ranking position), \textit{nDCG@1} ranges from $0.16$ in the $D8$ dataset to $0.74$ in $D1$ (and it is  above $0.41$ in half of the datasets). Besides, \textbf{TPDR} presents an increasing $nDCG$ when inspecting higher  ranking positions, demonstrating that the relevant items, when not retrieved immediately in the 1$^{st}$  or 2$^{nd}$ position of the rank, are usually placed up to 10$^{th}$   position. The $recall@10$ of $0.88$ on the largest dataset demonstrates \textbf{TPDR}'s ability to represent and retrieve products even when the number of queries is very large. In the worst scenarios, the $recall@10$ is $0.32$ ($D8$), where there exists a large  discrepancy between the query and product description terms, which is emphasized by the very low BM25  effectiveness in this dataset, that only reaches a $recall@10$ of $0.13$.\looseness=-1 

\vspace{-0.3cm}
\begin{figure}[htb]
\centering
\includegraphics[width=0.37\textwidth]{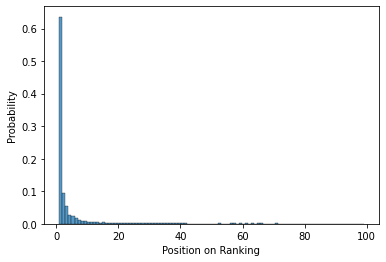}
\vspace{-0.3cm}
\caption{\small{Ranking Distribution of Product Description Position}}
\label{dist_pos_ranking}
\end{figure}
\vspace{-0.3cm}

Figure \ref{dist_pos_ranking} presents the number of standardized descriptions in each retrieved position for all datasets. Note that most SDs in the ranking distribution are present up to the 10$^{th}$ ranking position. Indeed, the largest portion of relevant SDs is found in the first positions of the rank (1$^{st}$ or 2$^{nd}$). With a few exceptions, all relevant standardized products can be found by examining the descriptions up to the 100th position. 
Considering our proposed model \textbf{TPDR}, we noticed that 71\% of the standardized descriptions are retrieved up to the 5$^{th}$ position of the ranking and 85\% of actual descriptions are up to the 100$^{th}$ position.\looseness=-1

When compared to the purely semantic model \textbf{TPDR (noRR)}, 44\% of the real products were retrieved up to the 5$^{th}$ position of the ranking with this model, while 72\% of actual descriptions were presented up to the 100$^{th}$ position. Finally, the purely syntactic model -- BM25 -- retrieved 22\% of the real products up to the 5$^{th}$ position of the ranking, with only 37\% of actual descriptions present up to the 100$^{th}$ position. In sum, there is a significant improvement in the results achieved by \textbf{TPDR} that combines both syntactic and semantic approaches  by means of a strong semantic ranker for producing a reduced and effective candidate list and a syntactic-based  re-ranking strategy to deal with product specificities.\looseness=-1

\newpage

\subsubsection{\textbf{Analysis of DP's Association}}~\label{results_DP}

In this section, we demonstrate the results regarding the association of DP's obtained by the proposed approach and the considered baselines. It is important to point out that these analyzes were performed considering only datasets D1 -- D9 (whose DP search space is composed of 3972 different DPs) since datasets D10 and D11 are associated with only one DP, namely Orthoses, Prostheses, and Medical Materials - \textbf{OPMM} and \textbf{Medicines}.\looseness=-1

In the previous section, we noticed that 71\% of the real standardized descriptions are present up to the 5$^{th}$ ranking position. Despite being very effective, this result still leaves room for improvement as 29\% of standardized descriptions are not retrieved in the top five positions. For this reason, the Correct Association of DP's is critical in cases where the standardized description is not retrieved or is retrieved very far in the rank, since it  facilitates the search (e.g., filtered by DP) for the  actual associated SD.\looseness=-1

Considering  \textbf{TPDR}, in our evaluation, we observed that, even when the  standardized description is not retrieved at the top of the rank, the correct DP is present in the $1^{st}$ position  in 80.2\% of the queries. Furthermore, the associated correct DP is retrieved up to the 5$^{th}$ ranking position in 88.2\% of queries. \looseness=-1

With the semantic model without re-ranking -- \textbf{TPDR (noRR)} -- the correct DP was present in the first ranking position in only 61.5\% of queries and among the top five positions in just 74.5\% of cases. Finally, the BM25 syntactic model returned the correct DP in the first position of the ranking in only 38.1\% of queries and among the top five positions in 48.5\% of cases. These results reinforce the need for combining syntactic and semantic information to adequately deal with the posed  problem.\looseness=-1

\subsubsection{\textbf{Manual Inspection of the Obtained Results}}

We discuss next some qualitative results of our  proposed approach \textbf{TPDR}.  We present in Table \ref{tab:exemplosprimeiraposicao} examples of descriptions retrieved in the $1^{st}$  ranking position. In none of the cases, purely syntactic approaches would be able to retrieve the description correctly since, considering the client and standardized descriptions, both are composed of different words with similar  meanings. There are also abbreviations, syntactically different names, or even typographical errors in all descriptions. Therefore, our current proposal manages to overcome the syntactic limitations, achieving semantically superior results.\looseness=-1

\vspace{-0.7cm}
\begin{table}[!h]
    \centering
        \begin{figure}[H]
        	\centering
\includegraphics[width=0.45\textwidth]{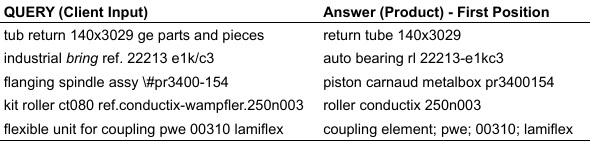}
        \end{figure}
    \vspace{-.4cm}
    \caption{Descriptions retrieved in the $1^{st}$  ranking position.}
    \label{tab:exemplosprimeiraposicao}
\end{table}
\vspace{-1cm}

From a pure semantic perspective, \textbf{TPDR} is similar to \textbf{TPDR(noRR)}. However, it is possible to notice that the ranking results improve in almost all cases in Table 2, when the syntactic re-ranker is applied, as is the case of the description "font semikron msk b 500/445-1.5 (9727)", seen in Table 4,  retrieved in the $5^{th}$ position by TPDR(noRR) and in $3^{rd}$ by the full TPDR solution.


Table \ref{tab:exemplos_n_posicao} presents descriptions retrieved erroneously in the $1^{st}$ position, but with the standardized product ranked up to the 3$^{rd}$ position. We can see that there is high similarity between them, often differing in aspects related to model, size, and other characteristics. \looseness=-1

\begin{table}[!ht]
    \centering
        \begin{figure}[H]
        	\centering
\includegraphics[width=0.49\textwidth]{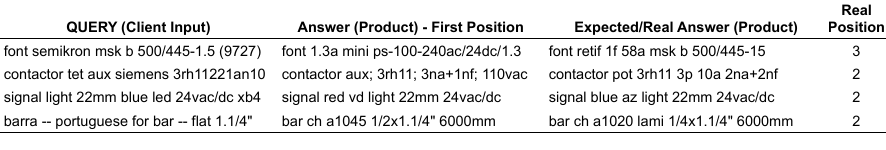}
        \end{figure}
    \vspace{-.3cm}
    \caption{Descriptions retrieved erroneously in the $1^{st}$ position, with the correct SD present in the top 3.}
\vspace{-0.5cm}
    \label{tab:exemplos_n_posicao}
\end{table}

Finally, in Table \ref{tab:exemplos_erros}, we present descriptions retrieved erroneously but with the correct associated DP in the $1^{st}$ position of the rank. 
Note that, as shown in the result of the previous table, most of the items retrieved erroneously are due to product specificities. However, the correct association of the PD helps to filter out the correct description, as it limits the search for similar products.\looseness=-1

\vspace{-.7cm}
\begin{table}[!h]
    \centering
        \begin{figure}[H]
        	\centering
\includegraphics[width=0.51\textwidth]{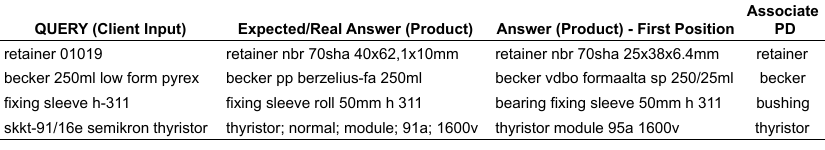}
        \end{figure}
    \vspace{-.3cm}
    \caption{Description retrieved erroneously in the top, but with PD correctly associated.}
    \label{tab:exemplos_erros}
\end{table}
\vspace{-.7cm}

The presented set of experiments so far demonstrated the flexibility and the proposal's effectiveness. All these results were properly reported and discussed with the contracting company -- Astrein -- which was very satisfied with the proposed solution.\looseness=-1
\section{Conclusion}\label{sec:con}

In this paper, we formalized the Product and Class Description Match problem -- matching an item described by a client with a product described in a catalog -- as a ranking task: given an initial client product specification (query), return the most appropriate standardized description (response). We demonstrated the complexity of this problem since the client's product descriptions are often noisy, short/uninformative, and may be formed of multi-language specifications. Also, the initial client specification (IS) may differ significantly from the standardized description (SD). To deal with this complex problem, we proposed \textbf{TPDR}, a two-step method that is able to explore the semantic correspondence between IS and SD by exploiting attention mechanisms, contrastive learning, and a syntactic-based re-ranking approach. 
In our evaluation, considering  11 datasets from a real company, covering different application contexts and dataset sizes, \textbf{TPDR} was able to retrieve the correct standardized product before the $5^{th}$ ranking position in \textbf{71\%} of the cases. Our proposed approach also managed to retrieve the correct Description Pattern in the first position of the ranking in \textbf{80.2\%} of the queries. Moreover, the effectiveness gains over the baselines (purely syntactic or semantic) reach up to 3.7 times, solving cases that none of the approaches in isolation could do. 

In future work, we plan to advance mainly in five directions: (i) evaluating other varieties of transformers as encoders~\cite{Cunha2023A,Cunha2023B}; (ii) further investing in fine-tuning the models due to the specificity of the vocabularies~\cite{AndradeB23,Cunha2020}; (iii)  evaluating alternative re-ranking score functions with dataset-specific coefficients; and, finally (iv)  investigating new strategies to increase the recall of the first retrieval step (Embedding-based Retrieval)~\cite{Bianco}, for instance by a semantic expansion of specifications~\cite{Belem,Viegas19,Viegas20}. 

\bibliographystyle{ACM-Reference-Format}
\bibliography{main}

\end{document}